\documentclass[11pt]{article} 
\pdfoutput=1
\usepackage{graphicx, amssymb, amsmath, geometry}
\graphicspath{{figs/}}

\newcommand{\figw}[2]{\includegraphics[width=#1\textwidth]{#2}}
\newcommand{\figh}[2]{\includegraphics[height=#1\textheight]{#2}}
\newcommand{\me}{\mbox{Mini-EUSO}}
\newcommand{\je}{\mbox{JEM-EUSO}}
\newcommand{\uhecrs}{\mbox{UHECRs}}
\newcommand{\tp}{\ensuremath{\mathrm{TP}}}
\newcommand{\tn}{\ensuremath{\mathrm{TN}}}
\newcommand{\fp}{\ensuremath{\mathrm{FP}}}
\newcommand{\fn}{\ensuremath{\mathrm{FN}}}

\begin{document}
\begin{center}\large\bfseries
	Neural Network Based Approach to Recognition of Meteor Tracks\\
	in the Mini-EUSO Telescope Data
\end{center}

\begin{center}
	Mikhail Zotov$^{1}$,
	Dmitry Anzhiganov$^{1,2}$,
	Aleksandr Kryazhenkov$^{1,2}$,
	Dario~Barghini$^{3,4,5}$,
	Matteo~Battisti$^3$,
	Alexander~Belov$^{1,6}$,
	Mario~Bertaina$^{3,4}$,
	Marta~Bianciotto$^{4}$,
	Francesca~Bisconti$^{3,7}$,
	Carl~Blaksley$^{8}$,
	Sylvie~Blin$^{9}$,
	Giorgio~Cambi{\`e}$^{7,10}$,
	Francesca~Capel$^{11,12}$,
	Marco~Casolino$^{7,8,10}$,
	Toshikazu~Ebisuzaki$^{8}$,
	Johannes~Eser$^{13}$,
	Francesco~Fenu$^{4,\dagger}$,
	Massimo~Alberto~Franceschi$^{14}$,
	Alessio~Golzio$^{3,4}$,
	Philippe~Gorodetzky$^{9}$, 
	Fumiyoshi~Kajino$^{15}$,
	Hiroshi~Kasuga$^{8}$,
	Pavel~Klimov$^{1,6}$,
	Massimiliano~Manfrin$^{3,4}$,
	Laura~Marcelli$^{7}$,
	Hiroko~Miyamoto$^{3}$,
	Alexey~Murashov$^{1,6}$,
	Tommaso~Napolitano$^{14}$,
	Hiroshi~Ohmori$^{8}$,
	Angela~Olinto$^{13}$,
	Etienne~Parizot$^{9,16}$,
	Piergiorgio~Picozza$^{7,10}$,
	Lech Wiktor~Piotrowski$^{17}$,
	Zbigniew~Plebaniak$^{3,4,18}$,
	Guillaume~Pr{\'e}v{\^o}t$^{9}$,
	Enzo~Reali$^{7,10}$,
	Marco~Ricci$^{14}$,
	Giulia~Romoli$^{7,10}$,
	Naoto~Sakaki$^{8}$,
	Kenji~Shinozaki$^{18}$,
	Christophe De~La~Taille$^{19}$,
	Yoshiyuki~Takizawa$^{8}$,
	Michal~Vr{\'a}bel$^{18}$ and Lawrence~Wiencke$^{20}$
\end{center}

\begin{flushleft}
	$^1$\quad Skobeltsyn Institute of Nuclear Physics, Lomonosov Moscow
		 State University, Moscow 119991\\
	$^2$\quad {Faculty of Computational Mathematics and Cybernetics,
		 Lomonosov Moscow State University, Moscow 119991, Russia}\\
	$^3$\quad INFN,
	Sezione di Torino, Via Pietro Giuria, 1-10125 Torino, Italy\\
	$^4$\quad Dipartimento di Fisica, Universit\`a di Torino, Via Pietro
	Giuria, 1-10125 Torino, Italy\\
	$^5$\quad INAF, Osservatorio Astrofisico di Torino, Via Osservatorio 20, 10025 Pino Torinese, Torino, Italy\\
	$^6$\quad Faculty of Physics, M.V. Lomonosov Moscow State University, Moscow 119991, Russia\\
	$^7$\quad INFN, Sezione di Roma Tor Vergata, Via della Ricerca
	Scientifica 1, 00133 Roma, Italy\\
	$^8$\quad RIKEN, 2-1 Hirosawa, Wako, Saitama, 351-0198, Japan\\
	$^9$\quad Universit\'e Paris Cit\'e, CNRS, AstroParticule et
	Cosmologie, F-75013 Paris, France\\
	$^{10}$\quad Dipartimento di Fisica, Universita degli Studi di Roma Tor Vergata, Via della Ricerca Scientifica 1, 00133 Roma, Italy\\
	$^{11}$\quad Max Planck Institute for Physics, Föhringer Ring 6
	D-80805 Munich, Germany\\
	$^{12}$\quad KTH Royal Institute of Technology, SE-100 44 Stockholm, Sweden\\
	$^{13}$\quad Department of Astronomy and Astrophysics, The University
	of Chicago, IL 60637, USA\\
	$^{14}$\quad INFN - Laboratori Nazionali di Frascati, 00044 Frascati
	(Roma), Italy\\
	$^{15}$\quad Department of Physics, Konan University,	Kobe 658-8501, Japan\\
	$^{16}$\quad Institut Universitaire de France (IUF), 75231 Paris Cedex 05, France\\
	$^{17}$\quad Faculty of Physics, University of Warsaw, 02-093 Warsaw, Poland\\
	$^{18}$\quad National Centre for Nuclear Research, ul. Pasteura 7, PL-02-093Warsaw, Poland\\
	$^{19}$\quad Omega, Ecole Polytechnique, CNRS/IN2P3, 91128 Palaiseau, France\\
	$^{20}$\quad Department of Physics, Colorado School of Mines, Golden,
	CO 80401, USA\\
	$^\dagger$\quad Current address:
	Agenzia Spaziale Italiana, Via del Politecnico, 00133 Roma, Italy.
\end{flushleft}

\centerline{\textbf{Abstract}}
{\narrower
Mini-EUSO is a wide-angle fluorescence telescope that registers
ultraviolet (UV) radiation in the nocturnal atmosphere of Earth from the
International Space Station. Meteors are among multiple phenomena that
manifest themselves not only in the visible range but also in the UV. We
present two simple artificial neural networks that allow for recognizing
meteor signals in the Mini-EUSO data with high accuracy in terms of a
binary classification problem.  We expect that similar architectures can
be effectively used for signal recognition in other fluorescence
telescopes, regardless of the nature of the signal.  Due to their
simplicity, the networks can be implemented in onboard electronics of
future orbital or balloon experiments.
}


\section{Introduction}

The \je{} (Joint Exploratory Missions for Extreme Universe Space
Observatory) collaboration is developing a program of studying
ultra-high energy cosmic rays (\uhecrs) with a wide angle telescope from
a low Earth orbit~\cite{JEM-expastron, jemeuso-mission,
Bertaina:2021+i}.  The idea is based on the possibility to register
fluorescence and Cherenkov radiation in the ultraviolet (UV) range that
is emitted during development of extensive air showers generated by
primary particles hitting the atmosphere~\cite{Benson-Linsley-1981}.
There are several benefits of this technique in comparison with
ground-based experiments: (i)~it can provide a huge exposure necessary
for collecting sufficient statistics of these extremely rare events;
(ii)~the celestial sphere can be observed almost uniformly, which is
important for anisotropy studies; and (iii)~the whole sky can be
observed with one instrument.

It became clear at early stages of the development of the \je{} program
that an orbital telescope aimed at studying \uhecrs{} can serve as a
tool for exploring other phenomena that manifest themselves in the UV
range in the nocturnal atmosphere of Earth~\cite{jemeuso-atmos}.  It was
demonstrated by TUS, the world's first orbital fluorescence telescope
aimed for testing the technique of studying \uhecrs{} from space, that
such an instrument can provide data on transient luminous events,
thunderstorm activity, meteors, anthropogenic illumination of different
kinds, and other types of signals~\cite{SSR2017, JCAP2017}.  In
particular, observations of meteors are considered as an important
branch of studies in the \je{} program~\cite{jemeuso-meteors,
ABDELLAOUI2017245}.

The \je{} program is being implemented in a number of steps aimed at
development and testing of different aspects of a full-blown orbital
experiment. In particular, laser shots were successfully registered by a
fluorescence telescope looking down on the atmosphere within the
EUSO-Balloon mission~\cite{jemeuso-balloon}. A wide program of studies
is being performed with the EUSO-TA experiment~\cite{euso-ta}. In
2018--2019, the \me{} (Multiwavelength Imaging New Instrument for the
EUSO) telescope was built by the \je{} collaboration. It was brought to
the International Space Station (ISS) on 22 August 2019, by the Soyuz
MS-14 vehicle and has been operated since then as a part of an agreement
between the Italian Space Agency (Agenzia Spaziale Italiana; ASI) and
Roscosmos (Russia)~\cite{2021ApJS..253...36B, Casolino:2021jO,
2023RLSFN.tmp...23M, 2023RSEnv.284k3336C}.  The EUSO-SPB2 stratospheric
balloon equipped with a fluorescence and Cherenkov telescopes made a
short flight from Wanaka, New Zealand, in May 2023~\cite{spb2-20,
spb2-22, Eser:2023Dw}. All these instruments are aimed to be pathfinders
and test beds for full-size orbital experiments like K-EUSO~\cite{keuso}
and POEMMA~\cite{poemma}.

Similar to the other projects of the \je{} collaboration, the \me{}
telescope is registering multiple types of UV emission taking place in
the nocturnal atmosphere of Earth, among them signals of meteors.  A
series of studies is dedicated to their search and
analysis~\cite{Dario-20, Dario-21}.  In the present paper, we continue
our earlier research aimed at developing a method of recognizing meteor
tracks in the \me{} data with neural networks~\cite{MZDS-2023}.  A
motivation for the study is the following.  A conventional approach to
finding signals of meteors in the \me{} data is time consuming and prone
to numerous false positives. Thus, it is interesting to figure out if an
approach based on machine learning (ML) and artificial neural networks
(ANNs) can demonstrate higher efficacy than the conventional one so that
both approaches complement each other.  If so, it is interesting to test
if results can be achieved with simple neural networks that can be
implemented in the forthcoming orbital experiments, which are unlikely
to have powerful onboard processors.  These results can also be useful
for recognizing tracks of extensive air showers in the future
experiments since such signals resemble shapes and kinematics similar to
those produced be meteors, though at completely different time scales.
Finally, in case of the successful development of an ANN-based pipeline
for recognizing meteor signals in the \me{} data, it can be applied for
a search of track-like signals of different nature, including those that
mimic extensive air showers.  In what follows, we present a pipeline
consisting of two basic neural networks that demonstrate high
performance and can be trained on an ordinary PC.  The work continues a
series of studies fulfilled within the \je{} collaboration on
application of machine learning and neural networks to analysis of data
of fluorescence telescopes~\cite{2019ICRC...36..456V, szakacs_etal-2019,
2022icrc.confE.405F, 2022JSSE....9...72M}.  We do not present any
results of applying the suggested method to data analysis since this
will be covered in detail in a dedicated paper.

\section{Mini-EUSO Experiment}

The main components of the \me{} telescope include two Fresnel lenses
and a focal surface (FS). The lenses have a diameter of 25~cm with the
focal distance of the optical system equal to 300~mm. The FS has a
square shape with 2304 pixels. It is built of \mbox{36 multi-anode}
photomultiplier tubes (MAPMTs) Hamamatsu R11265-M64 each consisting of
$8\times8$ pixels.  All MAPMTs are grouped into nine so-called
elementary cell (EC) units. Every EC unit has its own high-voltage
system, which operates independently of the others providing necessary
control of sensitivity of the respective MAPMTs.  A 2-mm thick UV filter
manufactured of the BG3 glass is located in front of each MAPMT. The
size of one pixel equals $2.88~\text{mm}\times2.88~\text{mm}$.  The
point spread function (PSF) has a size of $\sim$1.2 pixels.  \me{} has a
wide field of view (FoV) of $44^\circ\times44^\circ$ with spatial
resolution (FoV of one pixel) equal to
$6.3~\text{km}\times6.3~\text{km}$.  From the orbit of the ISS, an area
observed by the telescope exceeds $300~\text{km}\times300~\text{km}$.  A
detailed description of the instrument can be found
in~\cite{2021ApJS..253...36B}.

\me{} collects data in three modes. The D1 mode has a time resolution of
2.5~$\mu$s. This is called a D1 gate time unit (GTU). The D2 mode
records data integrated over \mbox{128 D1} GTUs. Finally, the D3 mode
operates with data integrated over $128\times128$ D1 GTUs resulting in
time resolution of 40.96~ms. To the contrary to the D1 and D2 modes, the
D3 mode does not have a trigger, and its data can be considered as a
series of videos with ``seasons'' corresponding to sessions of
observations and ``episodes'' corresponding to night segments of the ISS
orbit during a session. Each session takes around 12 h. With the orbital
period of the ISS equal to 92.9 min, a typical session includes eight
subsets of data taken during nocturnal segments of an orbit, with each
of them taking slightly longer than 1/3 of the period. Every  ``video''
has a resolution of $48\times48$ pixels and consists of $T/40.96~\mu$s
frames, where~$T$ is the duration of one nocturnal segment.
Observations are performed approximately twice per month through the
UV-transparent window at the Zvezda module with the schedule coordinated
with other experiments.  Due to this, background illumination varies
strongly from one session to another depending on the phase of the Moon
and the season.  The D3 mode allows for registering meteors and other
slow phenomena taking place in the night atmosphere of Earth. In what
follows, we use only data recorded during sessions 5--8 and 11--14 taken
from 19 November 2019 to 1 April 2020.  All artificial neural networks
discussed below were trained using data of seven sessions and tested on
the remaining session.  This way, we checked all possible combinations
of the eight sessions.

\section{Meteor and Background Signals}

Signals of meteors registered with \me{} have some features important
for the presented analysis:
\begin{itemize}
	\item A signal produced by a meteor in a pixel has a shape resembling
		the bell-like curve similar to the probability density function of
		the normal distribution.

	\item Meteor signals produce quasi-linear tracks in the focal
		surface.

	\item The number of hit (``active'') pixels in more than 75\% of
		meteor tracks is $\le$5, so that their footprints on the focal
		surface are small.

	\item Peaks of a meteor signal shift from one pixel to another
		(except for arrival directions close to nadir).

	\item
		There are multiple signals in the data with the shape similar to that
		of meteors but simultaneously illuminating large areas of the FS.

	\item Meteors are often registered on strong and quickly varying
		background illumination.

	\item The amplitude of a meteor signal is typically lower than amplitudes
		of some other signals in the FoV of \me{} registered simultaneously
		with the meteor.

	\item In some cases, it is impossible to judge unequivocally if a
		signal originated from a meteor or another source.
\end{itemize}

\begin{figure}[!ht]
	\centering
	{\figh{.22}{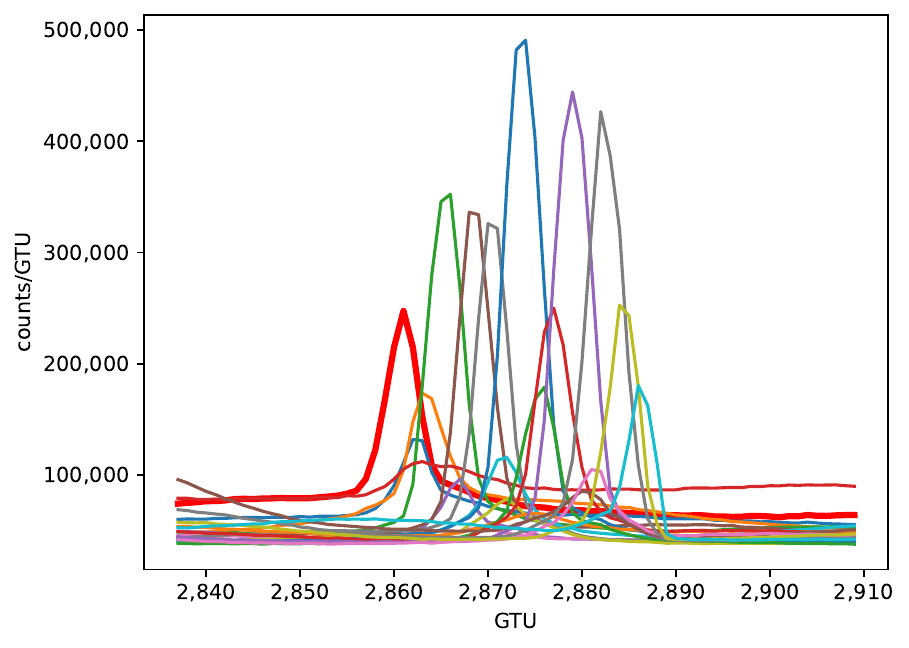}\quad\figh{.22}{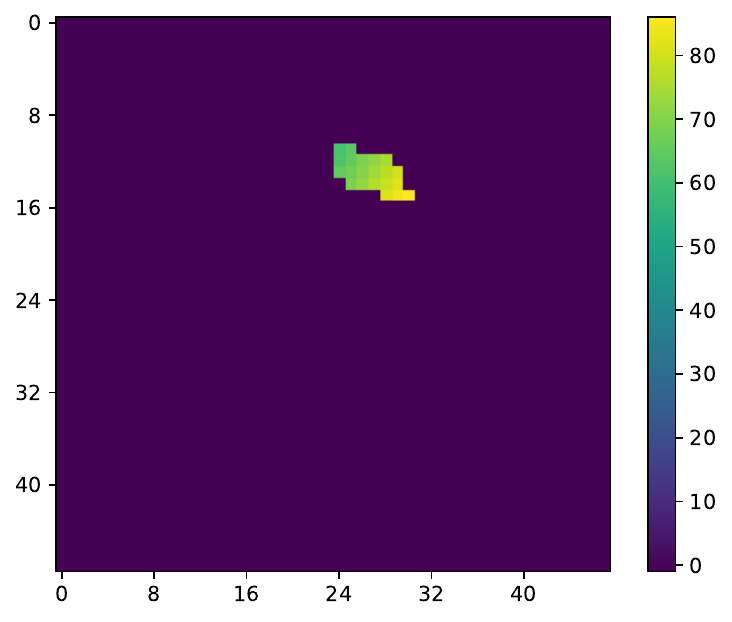}}
	{\figh{.23}{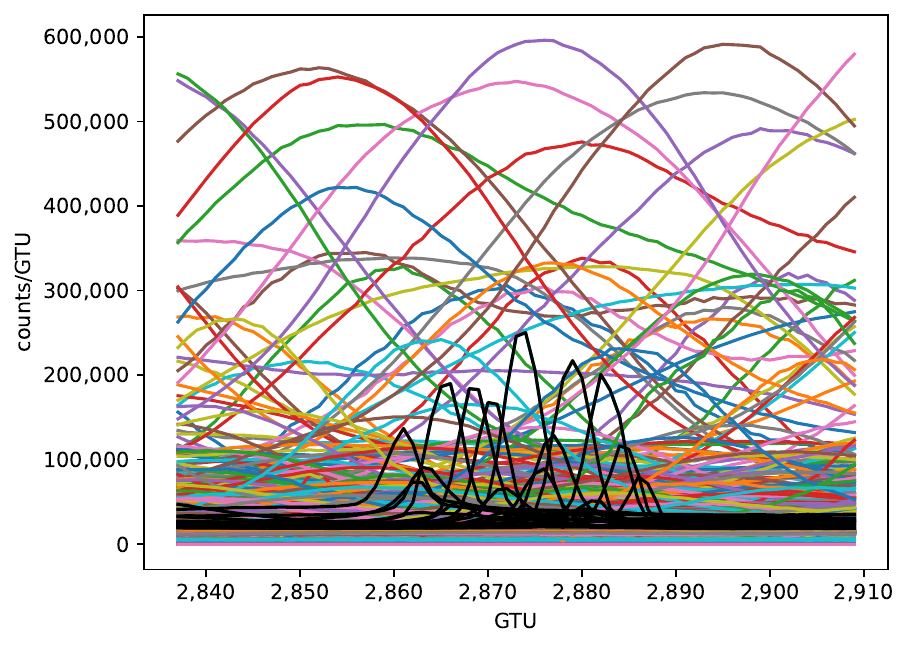}\quad\figh{.23}{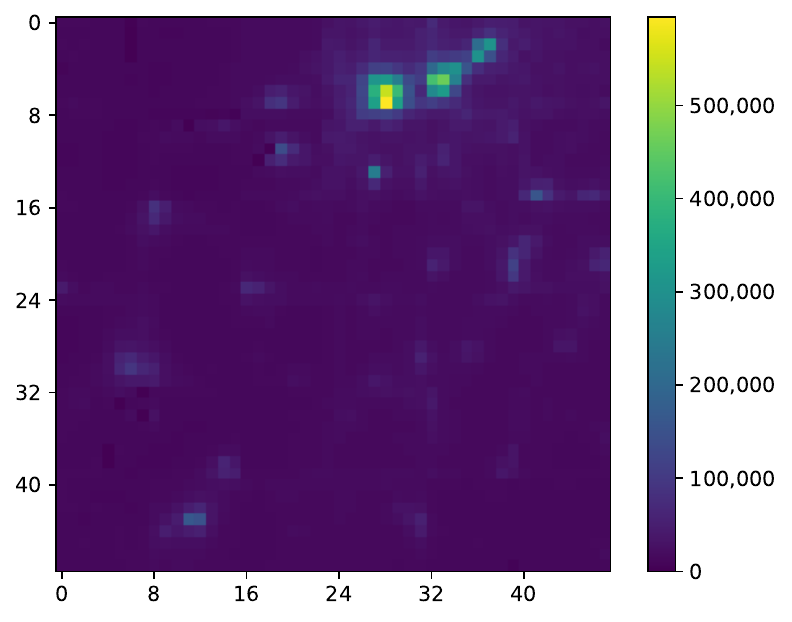}}
	\caption{An example of
	a clearly pronounced meteor signal registered by \me{}
	on 20 October 2019.
	({Top left}): signals in pixels that constitute the meteor signal.
	Signals in different pixels are shown in different colors.
	({Top right}): location of meteor pixels in the focal surface. Colors denote
	time shift of the peaks with respect to the first one (in units of D3 GTUs).
	({Bottom left}): all signals registered by \me{} simultaneously with the meteor.
	The black curves show the meteor signal.
	({Bottom right}): a snapshot of the focal surface made at the moment of
	maximum of the brightest meteor pixel (GTU 2874).}
	\label{fig:clear}
\end{figure}

\begin{figure}[!ht]
    
	\centering
	{\figh{.22}{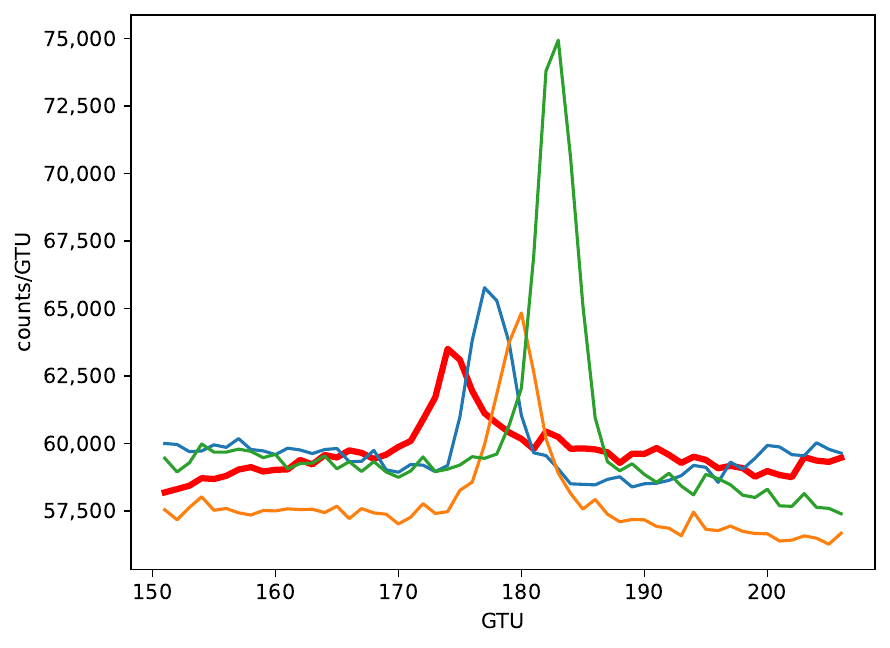}\quad\figh{.22}{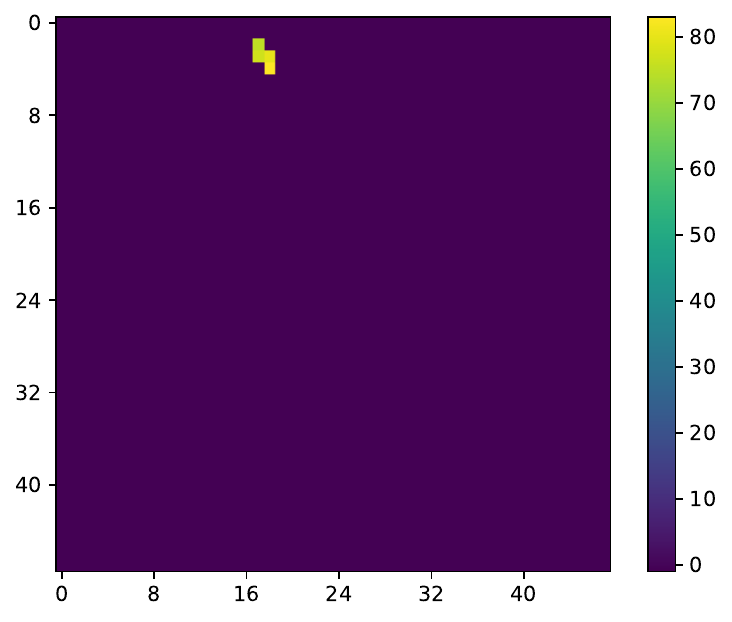}}
	{\figh{.22}{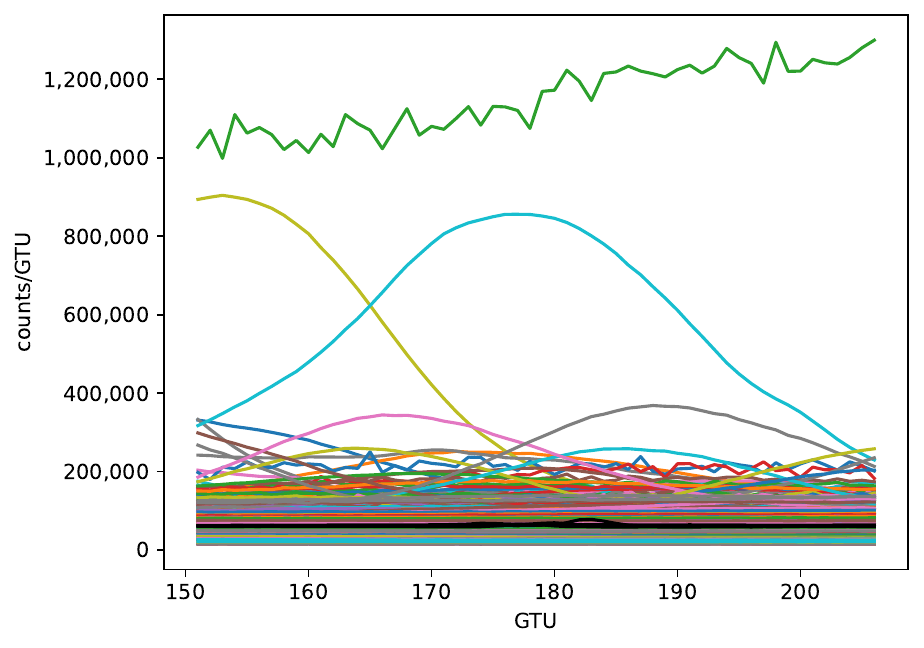}\quad\figh{.22}{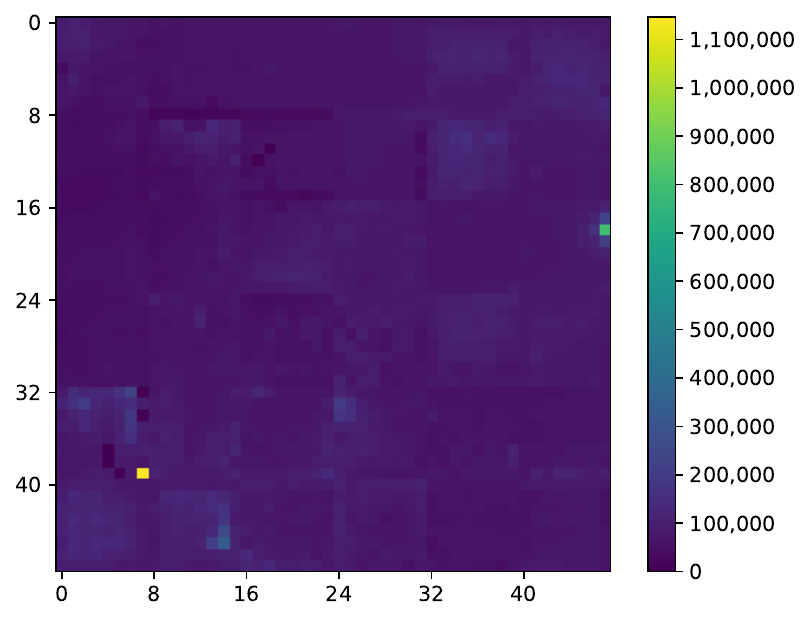}}
	\caption{A typical meteor
	signal registered by \me.
	({Top left}): signals in pixels that constitute the meteor signal.
	({Top right}): location of meteor pixels in the focal surface. Colors denote
	a time shift of the peaks with respect to the first one (in units of D3 GTUs).
	({Bottom left}): all signals registered by \me{} simultaneously with the meteor.
	The black curves show the meteor signal.
	({Bottom right}): a snapshot of the focal surface made at the moment of the
	brightest meteor signal (GTU 184).
	}
	\label{fig:typical}

\end{figure}

Let us discuss the most important of these features taking as an example
signals demonstrated in Figures~\ref{fig:clear} and~\ref{fig:typical}.

Figure~\ref{fig:clear} provides an example of a bright and clearly
pronounced meteor signal with numerous active pixels.  The top row shows
only the meteor signal, with the background illumination omitted.  It
can be seen that signals in every pixel have a typical shape resembling
the bell curve (see the left panel). The peaks are shifted in time with
respect to each other due to the meteor moving in the FoV of the
instrument, resulting in a quasi-linear track on the focal surface (see
the right panel).  The task of recognizing meteor signals might look
trivial after looking at these ``pure'' signals.  However, the FoV of
\me{} covers a huge area resulting in numerous different signals being
registered simultaneously, with many of them being much brighter than
those of meteors. This is demonstrated in the second row of
Figure~\ref{fig:clear}. The left panel shows shapes of all other signals
recorded simultaneously with the meteor with the meteor signal shown in
black. The right panel presents a snapshot of the FS made at the moment
of the maximum of the meteor signal. The brightest pixel of the meteor
has coordinates $(\text{row}, \text{column}) = (13, 27)$ and can be seen
as a small spot below a much brighter and extended area that appeared
due to anthropogenic illumination (sine-like curves in the left panel).
It is important to remark that the bottom rows of
Figures~\ref{fig:clear} and~\ref{fig:typical} demonstrate signals that
were flat-fielded during an offline analysis for the sake of clarity.

However, all results presented below were obtained using raw data as
they are recorded by the instrument.  This was performed in order to
understand how effectively is our method if implemented in onboard
electronics.

An example of a typical meteor is shown in Figure~\ref{fig:typical}.  It
has only four active pixels, and it is so dim in comparison with other
illumination registered simultaneously that it is hardly possible to
find it by eye in the bottom right panel presenting a snapshot of the
focal surface at the moment of the maximum brightness of the meteor.

A conventional way to find meteor signals in the \me{} or TUS data would
be to look for signals that can be fitted with the probability density
function of a Gaussian distribution or its sum with a polynomial in case
of non-stationary background illumination~\cite{Dario-20, Dario-21,
TUSmeteors}. The known range of possible speeds of meteors
(11--72~km~s$^{-1}$) together with information on the orbital speed of
the ISS ($\sim$7.7~km~s$^{-1}$), the FoV of one pixel, and the PSF size
allows one to estimate the variance of a Gaussian fit.  This also allows
for verifying the kinematics of a signal moving across the FS. The
latter step is of crucial importance since there are multiple signals in
the \me{} data that can be fitted with a Gaussian distribution but take
place simultaneously in big pixel groups without forming a track on the
focal surface.

The task of recognizing meteor patterns in the \me{} dataset with
machine learning methods can be considered as a binary classification
problem since one basically needs to separate meteor signals from all
the rest. Seemingly the most obvious way to tackle the task with
artificial neural networks is to employ supervised learning. Within this
approach, one can train an ANN using a labeled dataset. The dataset can
be either prepared by simulations or extracted from real experimental
data.  It is tempting to choose the first way since meteor signals
mostly have a bell curve shape similar to the density of the normal
distribution. However, realistic simulations are not trivial since the
background illumination is diverse, and sensitivity of different pixels
on the FS is not known.  This made us adopt the second approach.

We took two meteor datasets obtained by the \je{} collaboration and
complemented them by our own analysis to prepare a dataset suitable for
training and testing an ANN. It is necessary to stress once again that
the source of a considerable number of bell curve-like signals
registered with \me{} cannot be identified with confidence. Signals like
those shown in Figures~\ref{fig:clear} and~\ref{fig:typical} do not pose
a problem in this respect. However, the shape and kinematics of tracks
produced by meteors consisting of $\le$4 pixels are often confusing.
Another difficulty arises from dim signals on a strong and varying
background. As a result, it is impossible to obtain ground-truth labels
basing exclusively on the existing data set.  After several tests, we
confined the labeled dataset to signals the nature of which causes
little doubt. In particular, we excluded almost all signals occupying
two pixels. The resulting dataset used for training and testing ANNs
discussed below consisted of \mbox{1068 meteor} signals. Every record in
the dataset included a timestamp of a meteor, coordinates of active
pixels on the focal surface, and positions of the respective signal
peaks.

Since data obtained in the D3 mode do not have a trigger but are similar
to a series of videos each consisting of thousands of ``frames''
representing ``snapshots'' of the FS, there is a question how to extract
samples containing meteor and non-meteor signals for training and
testing datasets.  For example, one can create data chunks of size
$48\times48\times T$, where~$T$ is the number of time frames (D3 GTUs)
large enough to fit all meteors in the dataset and either center them on
meteor peaks or put them in a fixed position with respect to the
ibeginning of a meteor signal.  This will allow one to obtain a
``unified'' representation of meteor signals to an ANN thus simplifying
the task of their recognition. This way, non-meteor samples can be
extracted from the rest of the data in a random fashion.  However, this
is not the way the data flow can be analyzed onboard.  Besides this, the
above approach will leave us with mere 1068 meteor samples, which is not
sufficient to effectively train an ANN.  This made us use a sliding
window producing chunks that overlap by~$dt$~GTUs.  In what follows, we
present results obtained for $dt = 8$~GTUs that allowed us to avoid
losing short meteor tracks and provided reasonable accuracy of their
recognition. The procedure of labeling data chunks extracted this way
will be explained in detail in Section~\ref{sec:cnn}.

We split the task of meteor signal recognition into two steps.  First,
we trained an ANN to recognize three-dimensional data chunks that
contained meteor signals. After this, we employed another ANN to select
pixels containing respective signals. Each ANN thus solved a task of
binary classification. This allowed us to obtain lists of meteors
registered with \me{} together with their active pixels thus providing
information necessary for their subsequent analysis (reconstruction of
brightness, arrival directions etc.).

\section{Results}

An important question to discuss before presenting ANNs is how to
evaluate their performance.  It is usually advised to use balanced
datasets whenever possible, both for training and testing. In this case,
the Receiver Operating Characteristic Area Under the Curve (ROC AUC) is
one of the popular performance metrics~\cite{2006PaReL..27..861F}.
Recall that the ROC curve is a plot of the true positive rate against
the false positive rate at various threshold settings.  Given one
randomly selected positive instance and one randomly selected negative
instance, AUC is the probability that the classifier will be able to
tell which one is which. Due to its definition, ROC AUC does not depend
on the sample size.

However, the number of meteor signals in the \me{} data is negligibly
small in comparison with the number of non-meteor signals, so that using
balanced datasets for testing would provide unrealistic results, while
using them for training would not represent the full diversity of
non-meteor signals thus resulting in lower performance and numerous
false positives during tests.  Thus, we unavoidably arrive at the
necessity to use strongly imbalanced datasets both for training and
testing. It is argued in the literature that ROC AUC does not act as a
fully adequate performance metric in this case, and other metrics should
be used instead, see, e.g.,~\cite{2009PaReL..30...27F,
2015PLoSO..1018432S, ZHU202071}.  In what follows, we provide results
obtained in terms of three more metrics besides ROC AUC.  These are the
Precision--Recall (PR) AUC, the Matthews correlation coefficient (MCC),
and the $F_1$ score. One more metric will be introduced below.

Recall that the PR AUC equals an area under the plot of precision vs. recall with
\[
	\text{Precision} = \frac{\tp}{\tp + \fp}\,, \qquad
	\text{Recall} = \text{TPR} = \frac{\tp}{\tp + \fn}\,,
\]
where \tp, \fp, and \fn{} are the number of true positives, false
positives, and false negatives as classified by the model, respectively.
The Matthews correlation coefficient can be
calculated from the confusion matrix as
\[
	\text{MCC} = \frac{\tp\cdot\tn - \fp\cdot\fn}
					{\sqrt{(\tp+\fp)(\tp+\fn)(\tn+\fp)(\tn+\fn)}}\,,
\]
where \tn{} is the number of true negatives. Finally, the $F_1$ score is
the harmonic mean of the precision and recall. It can be presented as
\[
	F_1 = \frac{2\tp}{2\tp+\fp+\fn}.
\]

Notice that these metrics will be applied to three-dimensional data
chunks that partially overlap due to the employment of a sliding window
for preparing input datasets. This can lead to a situation when a part
of chunks containing a meteor signal are classified as non-meteor ones
while others are classified properly, so that the value of a performance
metric might be misleading.  Since we are interested in maximizing the
number of recognized meteors but not meteor chunks, it can be beneficial
to also introduce a metric expressed in terms of the original 1068
meteors.  Probably the most straightforward way is to use 1 $-$ FNR
(met), where FNR (met) is the false negative rate of meteor signals
represented as the number of meteors lost by the classifier divided by
the total number of meteors in a session used for testing.

All these metrics are equal to~1 for a perfect model. The MCC equals
$-1$ for the worst possible model; other metrics give~0 in this case.

\subsection{Recognition of Meteor Data Samples}
\label{sec:cnn}

One of the crucial questions to be solved before training an ANN is how
to prepare and organize input data. In our case, the question is
twofold. We need to decide how to label three-dimensional chunks into
those that contain a signal of a meteor and those that do not. Besides
this, we need to choose the size of data chunks $P\times P\times T$,
where $P\times P$ defines the size of a square on the focal surface
(measured in pixels), and~$T$ is the number of time frames
(``snapshots'' of the FS).

A data chunk was labeled as containing a meteor signal in case there
were at least \mbox{two meteor} pixels inside a $P\times P$ area with
peaks of the signals located within~$T$ GTUs.  The reason is that we do
not have a straightforward way to decide if a signal originates from a
meteor with just a single active pixel. On the other hand, putting a
more strict cut on the number of active pixels inside a chunk ($\ge$3)
results in a loss of short meteor tracks that have only two active
pixels.

As for the number of time frames in data chunks, we tested $T=8, 16, 32,
48, 64$, and 128, which covers the range of duration of meteor signals
in the dataset. Values $T=32$, 48, and 64 demonstrated the best results
in our tests in terms of all metrics mentioned above with different
combinations of training and testing sessions regardless of the value
of~$P$, with $T=48$ showing in average marginally better performance
than the other two values.  This value is used in all figures and tables
presented below.

\begin{figure}[!ht]
	\centering
	{\figw{.55}{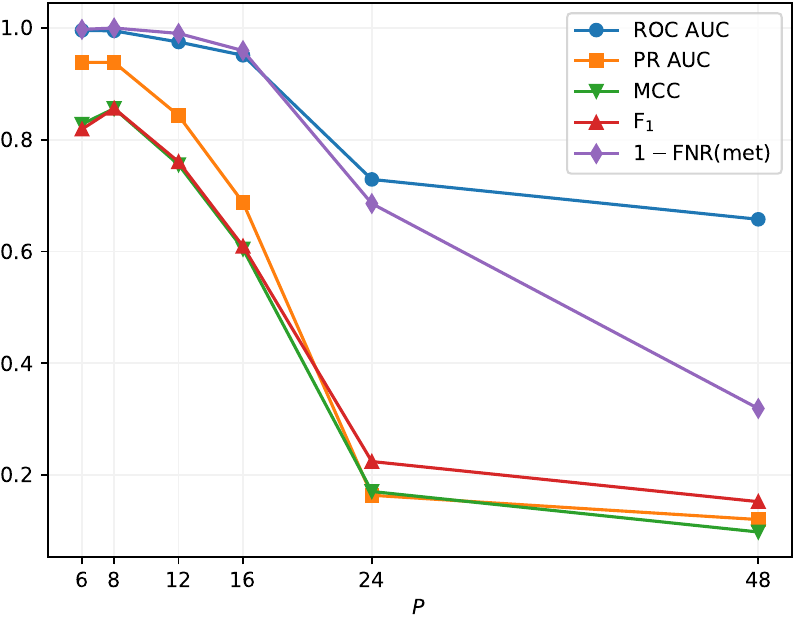}}
	\caption{{Mean
	values of different performance metrics as a function of
	the data chunk size~$P$ for models trained on all possible
	combinations of seven sessions of observations and tested on
	the remaining session. See the text for details.}}
	\label{fig:metrics}
\end{figure}

In~\cite{universe}, a simple convolutional neural network (CNN) was
employed to perform binary classification of two types of signals
registered with the TUS telescope. The instrument had a focal surface of
$16\times16$ pixels, and data arranged in $16\times16\times T$ chunks
worked well.  Thus, we first tried training a CNN for \me{} with data
chunks of the size $48\times48\times T$.  The input data was
standardized according to the formula $(X_i - \langle
X_i\rangle)/\sigma(X_i)$, where~$X_i$ is the signal in pixel~$i$,
$\langle X_i\rangle$ and $\sigma(X_i)$ are estimations of the mean and
the standard deviation during~$T$ time frames. However, as it was
briefly reported in~\cite{MZDS-2023}, this approach did not allow us to
obtain acceptable results. We tested numerous configurations of CNNs and
long short-term memory networks but failed to obtain ROC AUC $>0.75$ on
testing datasets.  A simple solution was found by splitting the FS into
smaller squares. We tested splitting with $P=24, 16, 12, 8$, and 6. In
order to avoid losing signals around the boundaries of these small
areas, we used overlapping by $P/2$ pixels in both directions.
Figure~\ref{fig:metrics} shows the behavior of mean values of different
performance metrics for varying~$P$ with models trained on all possible
combinations of seven sessions and tested on the remaining one.  In this
case, the same architecture of a CNN was used for all tests.  It can be
seen that performance expressed in terms of any metric increases quickly
for $P<24$. The PR AUC, the MCC, and the $F_1$ score change in a very
similar fashion while the values of the ROC AUC are close to those of
$1-$ FNR (met) for small~$P$. The best performance is reached for $P=8$
with the MCC and $F_1$ metrics slightly decreasing for $P=6$.  Thus,
data chunks of the size $8\times8\times48$ are used in what follows.

\begin{figure}[!ht]
	\centering
	{\figw{.4}{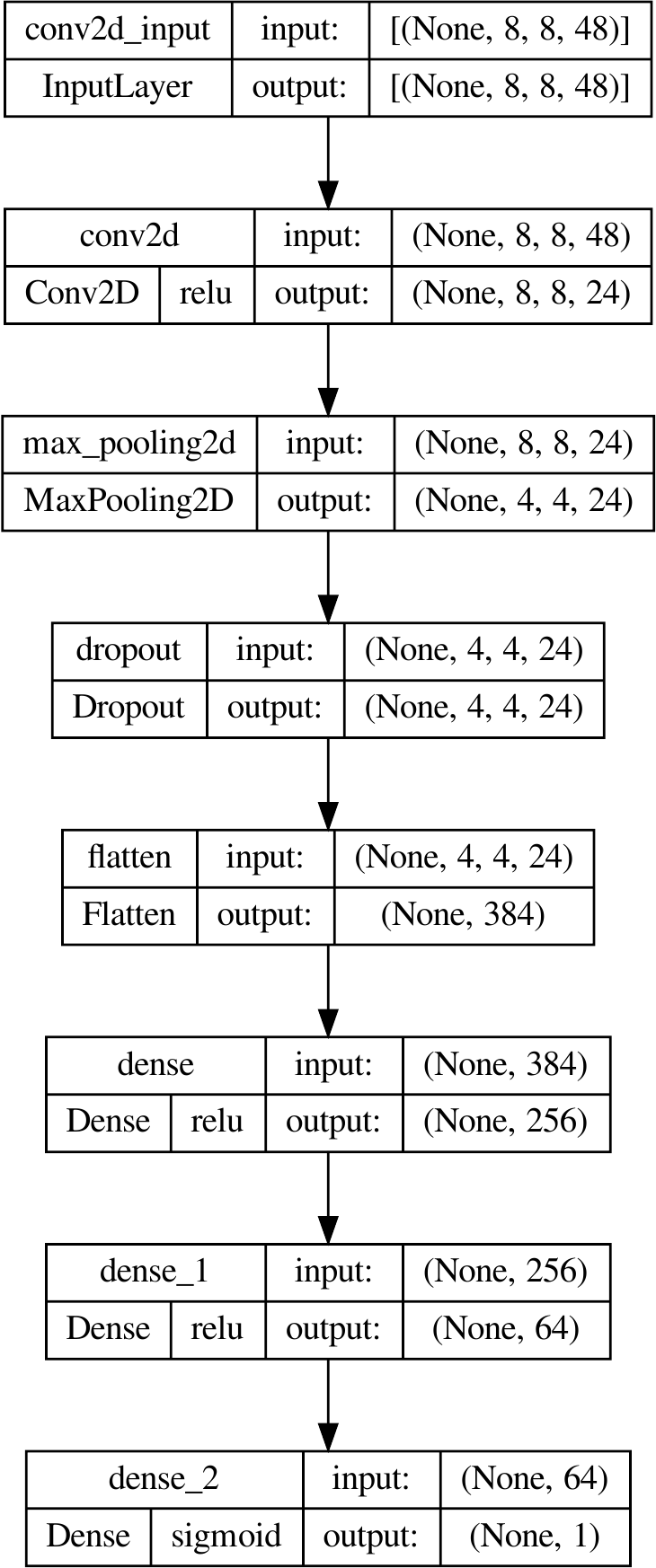}}

	\caption{Architecture of the CNN used for binary classification in
	meteor and non-meteor data chunks. The total number of trainable
	parameters equals 125,465.}

	\label{fig:cnn}
\end{figure}

A CNN that we adopted for classifying meteor chunks consists of a
convolutional layer with 24 filters and a kernel of size~3.  It utilizes
ReLU as an activation function and the L2 kernel regularizer with
factor~0.1. The convolutional layer was followed by a maxpooling and
dropout layers and two fully connected layers with 256 and then 64
neurons. Adam was used as an optimization algorithm. Sigmoid was
employed as an activation function in the output layer. The architecture
is shown in Figure~\ref{fig:cnn}.

The way we employed to prepare input data allowed us to obtain
$\approx$18,000--24,000 thousand meteor chunks of the size
$8\times8\times48$ depending on the set of sessions used for training,
see details in Table~\ref{tab:cnn_data}.  These chunks were augmented
then by the standard procedure of image rotation thus providing four
times more samples.  Non-meteor data chunks were selected in a random
fashion, with their number being six times larger than the total number
of meteor chunks (after augmentation).  Twenty percent of the training
dataset were used for validation during training. PR AUC was utilized as
a performance metric during the training process.  The loss function was
defined as binary crossentropy.  Validation loss was employed to adjust
the learning rate and to avoid overfitting.  Testing datasets included
100,000 non-meteor samples and all meteor data chunks available for the
particular session varying from 474 chunks for session~7 up to 6446
chunks for session~6, see Table~\ref{tab:cnn_data}.

\begin{table}[!ht]
	\caption{The top row:
	sessions of observations used for testing CNNs
	trained on data of all other sessions.
	The second row: the total number of $8\times8\times48$ chunks (after
	augmentation) with signals of meteors used for training the respective CNNs.
	The number of original chunks is four times less.
	The last two rows: the number of chunks with meteor signals and the
	real number of meteors, respectively, in test sessions.}
	\medskip
	\centering
	\begin{tabular}{lcccccccc}
		\hline
		{Test session} &{5}&{6}&{7}&{8}&{11}&{12}&{13}&{14}\\
		\hline
		Training meteor chunks&91,060 &72,124 &95,924 &81,188 &80,724 &88,456 &89,228 &86,820\\
		\hline
		Testing meteor chunks &1712  &6446  &474   &4180  &4274  &2341  &2148  &2772\\
		Testing meteors       &65    &280   &18    &186   &193   &106   &90    &130\\
		\hline
	\end{tabular}
	\label{tab:cnn_data}
\end{table}

Table~\ref{tab:cnn} contains values of five performance metrics
described above for different combinations of testing and training
datasets in the task of classification of $8\times8\times48$ chunks into
meteor and non-meteor groups.  For example, the column labeled as ``5''
presents results obtained with sessions 6--14 employed for training, and
session~5 used for testing the CNN.  It can be seen that zero out of
1068 meteors were lost in all sessions.  Notice however that values of
the PR AUC, the MCC, and the $F_1$ score vary considerably from one
session to another.

\begin{table}[!ht]
	\caption{Performance of the CNN on different sessions of
	observations. See the text for details.}
	\medskip
	\centering
	\begin{tabular}{lcccccccc}
		\hline
		{Test session} &{5}     &{6}     &{7}     &{8} &{11}    &{12}    &{13}    &{14}\\
		\hline
		ROC AUC      &0.992 &0.994 &0.999 &0.993 &0.994 &0.998 &0.993 &0.996\\
		PR AUC       &0.937 &0.955 &0.876 &0.933 &0.946 &0.973 &0.931 &0.956\\
		MCC          &0.872 &0.894 &0.732 &0.857 &0.888 &0.921 &0.782 &0.901\\
		F1           &0.873 &0.901 &0.718 &0.863 &0.892 &0.922 &0.776 &0.904\\
		\hline
		FNR (met)    &0     &0     &0     &0     &0     &0     &0     &0    \\
		\hline
	\end{tabular}
	\label{tab:cnn}
\end{table}

\subsection{Active Pixel Selection}

At the second step, we want to separate pixels of 3-dimensional meteor
chunks selected by the CNN into two groups: those containing the signal
of a meteor (active pixels) and all the rest.  Since meteor signals have
a typical shape resembling the bell curve, as shown in the top left
panels of Figures~\ref{fig:clear} and~\ref{fig:typical}, it is
straightforward to train a multilayer perceptron (MLP) to solve the
task.  The input dataset consists of vectors of length $T=48$ now.

We followed the same way of training and testing the neural network as
was used at the first stage. Namely, we trained an MLP on data extracted
from seven sessions and tested it on the remaining one with all possible
combinations of sessions. In order to avoid duplicate entries in the
training dataset, we extracted data vectors from chunks of the size
$48\times48\times48$. Due to a comparatively small shift~$dt=8$ GTUs, we
obtained samples with meteor signal peaks located at almost all possible
positions along the time axis.  The number of vectors (pixels)
containing meteor signals in training data sets was up to 30 thousand,
with non-meteor samples ten times greater.  The input data was
standardized similar to the CNN case.  Twenty percent of training
samples were utilized for validation.  Binary crossentropy was used as
the loss function and validation loss was employed to adjust the
learning rate and to avoid overfitting.  Testing was performed on all
vectors extracted from meteor chunks of the size $8\times8\times48$
selected by the CNN. The number of chunks used for testing of each
particular session data can be found in Table~\ref{tab:cnn_data}.

We compared a number of possible configurations of simple MLPs with one,
two, and three hidden layers and different number of neurons.  An
optimizer, activation functions and a performance metric used for
training were the same as for the CNN described above.
Table~\ref{tab:mlp} presents results obtained with a two-layer MLP with
96 and 64 neurons in the two layers, respectively.

\begin{table}[!ht]
	\caption{Performance of the MLP on different sessions of
	observations. See the text for details.}
	\medskip
	\centering
	\begin{tabular}{lcccccccc}
		\hline
		{Test session} &{5}     &{6}     &{7}     &{8}     &{11}    &{12}    &{13}    &{14}\\
		\hline                                                        
		ROC AUC      &0.992 &0.995 &0.993 &0.994 &0.996 &0.993 &0.995 &0.995\\
		PR AUC       &0.899 &0.932 &0.877 &0.916 &0.932 &0.887 &0.936 &0.928\\
		MCC          &0.790 &0.841 &0.744 &0.826 &0.847 &0.812 &0.809 &0.835\\
		F1           &0.794 &0.845 &0.737 &0.832 &0.852 &0.814 &0.810 &0.840\\
		Mean IoU     &0.808 &0.853 &0.773 &0.843 &0.861 &0.829 &0.825 &0.850\\
		\hline                                                        
		FNR (pxl) 	 &2/422 &2/1428&0/80  &7/928 &5/958 &0/492 &0/457 &2/630\\
		\hline
	\end{tabular}
	\label{tab:mlp}
\end{table}

Similar to Table~\ref{tab:cnn}, Table \ref{tab:mlp} presents results
illustrating performance of the MLP trained and tested on different
sessions of data collection.  Values of one more performance metric are
shown here, namely, the mean values of the intersection-over-union (IoU)
score. This function and its versions are often used in tasks of
labeling pixels of images.  It can be written as
\[
	\text{IoU} = \frac{\tp}{\tp + \fp + \fn}.
\]

It can be seen that values of the mean IoU metric are slightly above
those of the $F_1$ score.  The last line of the table shows the false
negative rate calculated for pixels containing meteor signals. Two
things can be easily noticed. First, the MLP did not properly recognize
18 out of 5395 active pixels, i.e., FNR (pxl)  $\approx0.33\%$. In this
sense the accuracy can be estimated as 99.67\% in average. On the other
hand, the worst result was obtained for the test run on data from
session~8 with FNR (pxl)  $\approx0.75\%$ so that the accuracy for this
particular session equals 99.25\%.

We have tried to address the same task with a few other machine learning
methods, among them logistic regression, K~nearest neighbors, the random
forest, and XGBoost. We have failed to outperform results demonstrated
with the MLP with any of them.

\section{Discussion}

We have demonstrated that a pipeline made of two simple neural networks,
a CNN and an MLP, can be used to effectively recognize meteor tracks in
the data of the \me{} fluorescence telescope that is observing the
nocturnal atmosphere of Earth in the UV band from the International
Space Station.  The CNN used to select three-dimensional data chunks
containing meteor signals managed to recognize properly all 1068 meteors
in the dataset.  The MLP employed to recognize pixels with meteor
signals in data chunks picked up by the CNN, reached an accuracy beyond
99\%. Neither of the ANNs puts high demands on computer resources
necessary for their training. Besides this, they perform surprisingly
fast during the classification stage with a major part of time being
spent on reading data from a data storage, thus strongly outperforming
the conventional algorithm.

We have seen in~\cite{universe} that an ANN trained on data with clearly
pronounced signals is able to identify patterns with low signal-to-noise
ratio. Such events are classified as false positives during tests but
their closer analysis reveals that their considerable part contains
``positive'' signals that were not found by the conventional algorithm
used to prepare training and testing datasets. A preliminary analysis of
signals marked as false positives in our tests has demonstrated that the
same situation takes place with meteor tracks, so that the list of
meteors can be extended with these newly found signals.  The same
applies to the list of active pixels.  This is an important advantage of
ML-based methods over conventional approaches.

One can anticipate that new experimental data might present new patterns
of non-meteor signals since observational conditions vary strongly from
one session to another. In this case, it might be necessary to extend
the training dataset and retrain the ANNs.  However, we do not expect
that the architecture of the CNN and the MLP will need to be modified
considerably.  On the other hand, it is clear that the presented ANNs
are not the only way of recognizing meteor tracks in the \me{} data
using methods of machine learning.  Still, it might be not easy to
exceed the accuracy of the suggested pipeline with simple models. We
plan to analyze other possible approaches, especially for the
segmentation part.  We are also going to address the task of solving the
same problem in one step, without splitting it into two parts.

Finally, it is worth mentioning that the presented method is not
confined to recognizing meteor tracks. Preliminary results of applying
the same approach and even the same trained models for recognizing
signals that mimic illumination expected from extensive air
showers born by ultra-high energy cosmic rays, are quite promising and
will be reported elsewhere.  We also expect that this pipeline
or a similar one can be implemented in onboard electronics of future
orbital missions to act as a trigger for track-like signals of different
nature manifesting themselves at various time scales.


\section*{Acknowledgments}
All neural networks discussed in the paper were implemented in Python
with TensorFlow~\cite{tf} and scikit-learn~\cite{sklearn} software
libraries.
The research of M.Z., D.A., and A.K. was funded by grant number 22-22-00367
of the Russian Science Foundation
(https://rscf.ru/project/22-22-00367/).

\section*{Abbreviations}
\begin{tabular}{ll}
	ANN & Artificial neural network \\
	AUC & Area under the curve \\
	CNN & Convolutional neural network \\
	EC  & Elementary cell \\
	FoV & Field of view \\
	FS  & Focal surface\\
	ISS & International space station \\
	MAPMT & Multi-anode photomultiplier\\
	MCC & Matthews correlation coefficient\\
	MLP & Multi-layer perceptron \\
	PR  & Precision-recall \\
	PSF & Point spread function \\
	ROC & Receiver operating characteristic\\
	UHECR & Ultra-high energy cosmic ray\\
	UV  & ultraviolet
\end{tabular}

\bibliographystyle{hieeetr}
\bibliography{mlmeteors}
\end{document}